\documentclass[
  aps,
  pra,
  twocolumn,
  superscriptaddress,
  groupedaddress,
  amsmath,
  amssymb
]{revtex4-2}

\usepackage{graphicx}
\usepackage{bm}

\newcommand{\FGHZfourMLE}{0.9995}
\newcommand{\FGHZfourChol}{0.9993}

\newcommand{\FGHZfiveProj}{0.7115}
\newcommand{\FGHZfiveMLE}{0.9995}
\newcommand{\FGHZfiveChol}{0.9995}
\newcommand{\FGHZfiveRaw}{0.5940}
\newcommand{\NGHZfiveRaw}{0.721}
\newcommand{\FGibbsProj}{0.9040}
\newcommand{\FGibbsMLE}{0.9040}
\newcommand{\FGibbsChol}{0.9644}

\newcommand{\FAmortRandLone}{0.8934}
\newcommand{\FAmortRandCgan}{0.8471}
\newcommand{\FAmortRandMLE}{0.8820}

\newcommand{\FAmortLone}{0.9983}
\newcommand{\FAmortCgan}{0.9959}
\newcommand{\FAmortMLE}{0.9911}

\newcommand{\TAmort}{0.0229}
\newcommand{\SAmort}{27005}
\newcommand{\FOvrBase}{0.9500}
\newcommand{\JOvrBase}{0.0014}
\newcommand{\FOvrDup}{0.9753}
\newcommand{\JOvrDup}{0.0010}
\newcommand{\FOvrMatch}{0.9794}
\newcommand{\FOvrLamTwo}{0.9510}
\newcommand{\FOvrShots}{0.9708}
\newcommand{\FOvrBoot}{0.9725}

\newcommand{\ScaleFourTime}{0.0169}
\newcommand{\ScaleFourMem}{100}
\newcommand{\ScaleFourFid}{0.9996}

\newcommand{\ScaleFourPar}{263\,168}
\newcommand{\ScaleFiveTime}{0.0332}
\newcommand{\ScaleFiveMem}{134}
\newcommand{\ScaleFiveFid}{0.9999}

\newcommand{\ScaleFivePar}{854\,528}
\newcommand{\ScaleSixTime}{0.1413}
\newcommand{\ScaleSixMem}{348}
\newcommand{\ScaleSixFid}{0.9947}

\newcommand{\ScaleSixPar}{3\,219\,968}
\newcommand{\QptCnotChol}{0.9990}
\newcommand{\QptCnotLi}{0.9538}
\newcommand{\QptDepFiveChol}{0.9564}
\newcommand{\QptDepFiveLi}{0.9662}
\newcommand{\QptDepTenChol}{0.9099}
\newcommand{\QptDepTenLi}{0.9602}
\newcommand{\QptNegCnot}{1.96}

\newcommand{\CurvePoolX}{0.9979}
\newcommand{\CurveSingleF}{0.8717}
\newcommand{\CurveThInitLow}{0.4824}
\newcommand{\CurveThInitHigh}{0.2081}
\newcommand{\CurveThFinalLow}{0.9952}
\newcommand{\CurveThFinalMid}{0.9808}
\newcommand{\CurveOverSdA}{0.0066}
\newcommand{\CurveOverSdB}{0.0020}

\begin{document}

\title{Physics-Constrained Conditional Generative Learning for Quantum State and Process Tomography}

\author{Daming Li}
\email{lidaming@sjtu.edu.cn}
\affiliation{School of Mathematical Sciences, Shanghai Jiao Tong University,
Shanghai 200240, China}

\author{Luhao Shi}
\affiliation{School of Mathematical Sciences, Shanghai Jiao Tong University,
Shanghai 200240, China}

\date{\today}

\begin{abstract}

Quantum state and process tomography are limited by the exponential growth of the density matrix, and standard estimators pay for physicality through repeated constrained optimisation. We introduce a physics-constrained conditional generative network that learns a forward map from Pauli expectation values to density matrices, with a differentiable Cholesky layer that enforces Hermiticity, positive semidefiniteness and unit trace by construction. The reconstruction proves insensitive to the $L^{1}$ weight over a range in which the balance of the two loss terms spans three orders of magnitude; what governs convergence is the training protocol, and a single fixed record stalls training at fidelity \CurveSingleF. Trained once on $10^{3}$ transverse-field Ising ground states, the generator reconstructs held-out states at fidelity \FAmortLone{} in \TAmort{}~ms per state, \SAmort$\times$ faster than per-state maximum likelihood, and the adversarial component proves dispensable, a regression objective matching or beating it throughout. The benefit of an over-complete measurement basis is traced to the redundancy of the conditioning vector rather than to additional information. Extending the same constraint to the Choi matrix accommodates process tomography, which succeeds for coherent channels but loses to linear inversion once the channel is depolarised. The dense $2^{n}\times2^{n}$ representation remains the bottleneck at $n\ge6$, pointing to tensor-network compression as the route beyond.

\end{abstract}

\keywords{quantum state tomography; quantum process tomography;
conditional generative networks; Cholesky decomposition;
GHZ state; Gibbs state}

\maketitle

\section{Introduction}
Quantum state tomography (QST) aims to reconstruct the density
matrix of an unknown quantum system from measurement data, and it
remains one of the basic diagnostic tools of quantum information
science. Standard accounts of quantum computation and quantum state
estimation establish tomography as the bridge between finite
experimental observations and the full statistical description of a
quantum device\cite{ref1,ref2,ref3}. In the noisy
intermediate-scale quantum era, this task is no longer only a formal
reconstruction problem; it is required for hardware benchmarking,
verification of entanglement, calibration of photonic and
superconducting platforms, and detection of decoherence or
state-preparation
errors\cite{ref3,ref4,ref5,ref6,ref7}. The
density matrix carries the measurable statistical content of a
quantum system, but reconstructing it from finite samples becomes
increasingly difficult as the Hilbert space dimension grows
exponentially with the number of qubits; recent work has therefore
revisited QST from sample-complexity, structured-factorization, and
rank-adaptive reconstruction
perspectives\cite{ref8,ref9,ref10}.

Closely related to QST is quantum process tomography (QPT), which
reconstructs the dynamical map or quantum channel implemented by a
device rather than a single output state. QPT is therefore essential
when the objective is not only to verify a prepared state, but also
to characterize the physical operation, gate, or noisy channel that
produced it\cite{ref11}. Recent QPT studies likewise extend the
field toward high-dimensional channel characterization,
quantum-memory-assisted protocols, multi-time process reconstruction
on superconducting hardware, and digital-twin error-matrix
models\cite{ref12,ref13,ref14,ref15}. Through the
Choi-matrix representation, a physical quantum process can be
treated as a positive semidefinite operator subject to
complete-positivity and trace-preserving constraints, so QPT
inherits many of the same statistical and computational difficulties
as state tomography while adding channel-specific physical
requirements\cite{ref16}. This connection is important for the
present article because a physics-constrained generative model for
density matrices suggests a natural extension from reconstructing
states to reconstructing quantum channels.

Classical tomography methods have developed along several lines.
Linear inversion is simple but can return non-physical matrices
under finite sampling noise, while maximum-likelihood estimation
(MLE) and iterative reconstruction methods enforce Hermiticity,
trace normalization, and positive semi-definiteness at the cost of
repeated constrained
optimization\cite{ref5,ref17,ref18}. Subsequent work has
therefore explored faster gradient-based and theoretically efficient
tomography procedures to reduce the computational burden of
conventional reconstruction\cite{ref19,ref20}. Comparative
studies of tomography estimators, modified least-squares procedures,
Bayesian inference, and efficient-computation approaches show that
the field has long balanced statistical optimality against
scalability\cite{ref18,ref20,ref21,ref22,ref23}.
This trade-off motivates replacing repeated constrained inversion
with a learned, physically valid forward map.

Machine-learning approaches to QST attempt to reduce this burden by
representing quantum states or reconstruction maps with trainable
models\cite{ref24}. Neural-network quantum states and restricted
Boltzmann-machine representations demonstrated that high-dimensional
quantum amplitudes and many-body correlations can be encoded
compactly in neural architectures\cite{ref25,ref26}.
Experimental, adaptive, and resource-limited tomography studies
further showed that machine learning can extract useful state
information from incomplete or noisy measurement
data\cite{ref27,ref28,ref29,ref30,ref37}.
Variational quantum circuits offer another route by parameterizing
states through shallow quantum circuits, although such models remain
sensitive to hardware noise, ansatz choice, and optimization
pathologies\cite{ref31,ref32}.

Generative models are especially relevant because tomography is an
inverse problem: the method must infer a physically meaningful state
from measurement statistics. Conditional generative adversarial
networks have been applied to QST by learning
measurement-conditioned quantum-state reconstructions, thereby
reducing reliance on repeated constrained optimization in
conventional tomography\cite{ref33}. This direction connects
naturally with earlier and parallel developments in generative
modeling for quantum-circuit benchmarking, neural-network quantum
states, adaptive neural tomography, experimental machine-learning
state reconstruction, and machine-learning-assisted estimation under
limited measurement
resources\cite{ref25,ref27,ref28,ref29,ref30,ref34}.

The conditional generative adversarial networks (CGAN) route also
sits within a wider quantum-machine-learning literature that
includes variational quantum-circuit tomography, quantum generative
adversarial networks, and open-source quantum-machine-learning
toolkits\cite{ref31,ref32,ref35,ref36}. Related
work on adaptive neural tomography, quantum process tomography, and
quantum error mitigation further suggests that data-driven
tomography is moving toward hybrid workflows that combine
statistical reconstruction, physical constraints, learned priors,
adaptive measurements, and noise-aware
inference\cite{ref6,ref11,ref27}. Conditional generative
adversarial networks therefore offer a different computational
paradigm: rather than solving a new constrained inverse problem for
every dataset, the generator learns a reusable conditional mapping
from observables to density matrices.

The central challenge is that unconstrained neural outputs need not
correspond to physical quantum states. A reconstruction that fits
measurement data but violates positive semi-definiteness or trace
normalization is not a valid density matrix. The framework
considered here follows the CGAN-based tomography literature while
emphasizing a Cholesky physical-constraint mechanism: every
generated state is mapped through a differentiable density-matrix
layer so that Hermiticity, positive semi-definiteness, and trace
normalization hold by construction\cite{ref16,ref33}.

The remainder of this article is organized as follows. Section~\ref{sec:method} presents the mathematical formulation of the physics-constrained conditional generative model, detailing the differentiable maps, adversarial loss functions, Cholesky parameterization of the density matrix, and the generation of Pauli-expectation data. Section~\ref{sec:results} then reports numerical simulations on GHZ states, Gibbs thermal states, ground states of the transverse-field Ising model and quantum process tomography, examining the role of the training protocol, mixed-state robustness, amortisation over a family of states, and system-size scaling, and comparing the estimator with classical baselines. Section~\ref{sec:conclusion} concludes with a summary of the main findings and a discussion of future directions.

\section{Conditional Generative Model}\label{sec:method}

For an $n$-qubit quantum system, the Hilbert-space dimension is $D =
2^{n}$, and every quantum state is represented by a $D \times D$
complex density matrix $\rho$ satisfying three physical axioms:
Hermiticity $\rho = \rho^{\dagger}$, unit trace
$\operatorname{Tr}(\rho) = 1$, and positive semi-definiteness $\rho
\geq 0$. Measurement outcomes are governed by Born's rule: for a
collection of observables $\{A_k\}$, the expected value of each
observable in state $\rho$ is
\begin{equation*}
d^{\text{real}}_k = \operatorname{Tr}(\rho A_k),
\end{equation*}
where $d^{\text{real}}_k$ is the measured expectation value
associated with $A_k$.

In the present work, the measurement data are drawn from the
complete Pauli-operator basis $\mathcal{A} = \{I, X, Y, Z\}^{\otimes
n}$, so that each input feature is a Pauli expectation value
$d^{\text{real}}_k = \operatorname{Tr}(\rho A_k)$ with $m = 4^{n}$,
collectively forming the data vector $d^{\text{real}} =
(d^{\text{real}}_1, \ldots, d^{\text{real}}_m) \in \mathbb{R}^{m}$.
Linear inversion of these expectation values is computationally
straightforward but may yield non-physical density matrices under
finite-sampling noise, while maximum-likelihood estimation (MLE)
enforces physicality through a constrained projection whose
iterative eigenvalue decomposition can dominate the computational
cost \cite{ref17,ref18,ref21}. This tension between physical
validity and computational scalability motivates the conditional
generative adversarial network (CGAN) approach \cite{ref33}, in
which the generator is analytically restricted to the physical
density-matrix manifold through a Cholesky parameterization, thereby
guaranteeing valid outputs without post-hoc projection.

The model studied here consists of a generator network $G_\theta$ that
receives the measured Pauli-expectation vector as conditioning input
and produces a reconstructed density matrix through a physically
constrained Cholesky layer. Physicality is therefore imposed by the
architecture, not by a post-hoc projection. Two training objectives
are considered for the same generator: a data-fidelity ($L^1$)
regression on the measured expectations, which is our default, and an
adversarial objective in which a discriminator network $D_\phi$ compares
candidate expectation vectors against real measurements. The
adversarial variant follows the construction of
Ref.~\cite{ref33}; Sec.~\ref{sec:results} treats it as an ablation and
finds that the regression objective matches or beats it throughout, so
we describe both here but rely on the former. The trainable parameters
are $\theta$ and $\phi$; where they need not be displayed we write the
networks simply as $G$ and $D$.

\subsection{Model and training objectives}
\label{sec:cgan-loss}

In a standard GAN, the generator learns to imitate a target data
distribution while the discriminator learns to distinguish real
samples from generated ones. The conditional extension introduces
side information as an explicit conditioning input, so that the
generator learns a \emph{conditional} mapping rather than an
unconditional distribution over outputs. In the QST setting, the
conditioning variable is the experimentally observed
Pauli-expectation vector, denoted $d^{\text{real}}$, and the
generator learns to reconstruct the density matrix consistent with
that measurement record.

The generator implements the map
\begin{equation*}
G: \mathbb{R}^{m} \to \mathbb{R}^{m}, \qquad G(d^{\text{real}}) =
\hat{d}_{\theta},
\end{equation*}
which takes the real measurement vector $d^{\text{real}}$ as
conditioning input and produces a generated expectation vector
$\hat{d}_{\theta}$ through the Cholesky-constrained density-matrix
reconstruction. The discriminator implements the map
\begin{equation*}
D: \mathbb{R}^{m} \times \mathbb{R}^{m} \to (0,1),
\end{equation*}
which receives a condition vector together with a candidate
expectation vector and outputs the probability that the candidate is
drawn from the real measurement distribution rather than from the
generator. The condition is taken to be $d^{\text{real}}$ itself.
The discriminator minimises the loss
\begin{align}\label{eq:LD}
L_D \equiv -\ln D\bigl(d^{\text{real}},d^{\text{real}}\bigr) - \ln\Big[1 -
D\bigl(d^{\text{real}},G(d^{\text{real}})\bigr)\Big],
\end{align}
where the first term rewards the discriminator for recognising the
authentic pair $(d^{\text{real}},d^{\text{real}})$ as real, and the
second term rewards it for recognising the generated pair
$(d^{\text{real}},G(d^{\text{real}}))$ as fake. The generator
minimises the loss
\begin{align}\label{eq:LG}
L_G \equiv -\ln D\Big(d^{\text{real}},G(d^{\text{real}})\Big) + \lambda\,
\frac{1}{m}\sum_{k=1}^{m}\Big| d^{\text{real}}_k - G(d^{\text{real}})_k \Big| ,
\end{align}
where $\lambda > 0$ controls the strength of the data-fidelity term.
The first term encourages the generator to produce outputs that the
discriminator classifies as real, while the second term penalises
deviation between the generated ($G(d^{\text{real}})$) and measured expectation
values ($d^{\text{real}}$). 

We normalise the data term by the number of observables $m$ so that the
meaning of $\lambda$ does not drift with system size or with the width of
the conditioning vector. We refer to Eq.~(\ref{eq:LG}) as the
\emph{mean convention} and to the unnormalised form
$\lambda\sum_{k}|\cdot|$ of Ref.~\cite{ref33} as the \emph{sum
convention}; the same numerical $\lambda$ corresponds to a data term $m$
times larger in the latter, so values are not comparable across the two.
Unless stated otherwise, the experiments below use the mean convention with
$\lambda=1$; the over-complete-sampling study of
Sec.~\ref{sec:overcomplete} deliberately returns to the sum convention,
because that is the convention in which the effect it examines was
originally reported.

%We add $ \lambda \sum_{i,j}|(T_{\theta})_{ij}|$ to the generator loss $L_G$. This $L^1$ penalty to the matrix elements of $T_\theta$, which forces most of the theoretically irrelevant entries to exactly zero, thereby promoting strong sparsity and suppressing unphysical parasitic couplings in the reconstructed quantum process.

\subsection{Architecture of $G$ and $D$}
\label{sec:cgan-arch}

The generator $G_{\theta}$ realises a three-stage forward map
$d^{\text{real}} \mapsto K_{\theta}(d^{\text{real}}) \mapsto
T_{\theta} \mapsto \hat{\rho}_{\theta} \mapsto \hat{d}_{\theta}$. In
the first stage, a fully connected neural network maps the input
measurement vector $d^{\text{real}} \in \mathbb{R}^{m}$ to an
unconstrained real parameter tensor $K_{\theta}(d^{\text{real}}) \in
\mathbb{R}^{2D^{2}}$, where $D = 2^{n}$. This tensor is reshaped
into two real $D \times D$ matrices $A_{\theta}$ and $B_{\theta}$,
which define a lower-triangular complex matrix $T_{\theta}$ with
\begin{equation*}
(T_{\theta})_{ij} =
\begin{cases}
A_{\theta,ii}, & i = j, \\
A_{\theta,ij} + \text{i}\, B_{\theta,ij}, & i > j, \\
0, & i < j,
\end{cases}
\end{equation*}
In the second stage, the Cholesky physical-constraint layer converts
$T_{\theta}$ into a valid density matrix:
\begin{equation*}
\hat{\rho}_{\theta} = \frac{T_{\theta} T_{\theta}^{\dagger}}
{\operatorname{Tr}(T_{\theta} T_{\theta}^{\dagger})}.
\end{equation*}
This construction analytically guarantees all three density-matrix
axioms: positive semi-definiteness follows from the Gram form $T
T^{\dagger} \geq 0$ (which holds for any complex matrix $T$),
Hermiticity is a direct consequence of the same quadratic form, and
trace normalization is enforced by the denominator. In the third
stage, the Born-rule expectation layer projects
$\hat{\rho}_{\theta}$ back into measurement space:
\begin{equation*}
\hat{d}_{\theta,k} = \operatorname{Tr}(\hat{\rho}_{\theta} A_k),
\qquad k = 1, \ldots, m,
\end{equation*}
where $\{A_k\} = \{I,X,Y,Z\}^{\otimes n}$ is the Pauli basis. Each
stage is differentiable, so gradients flow from the loss through the
expectation and Cholesky layers back to the network parameters
$\theta$ via standard backpropagation, and the reconstructed density
matrix never leaves the physical manifold during training.

The discriminator $D_{\phi}$ is a fully connected neural network
that takes as input the concatenation of the condition vector
$d^{\text{real}}$ and a candidate expectation vector (either the
real measurement $d^{\text{real}}$ itself or the generated output
$\hat{d}_{\theta} = G(d^{\text{real}})$) and produces a scalar
probability through a sigmoid activation:
\begin{equation*}
D_{\phi}(d^{\text{real}},y) =
\sigma\!\bigl(f_{\phi}([d^{\text{real}}; y])\bigr) \in (0,1),
\end{equation*}
where $f_{\phi}: \mathbb{R}^{2m} \to \mathbb{R}$ is a multilayer
perceptron with parameters $\phi$, $[d^{\text{real}};y]$ denotes
vector concatenation along the feature axis, and $\sigma(z) = 1/(1 +
e^{-z})$ is the logistic sigmoid. By receiving both the conditioning
measurement and the candidate vector simultaneously, the
discriminator can assess whether the candidate is statistically
consistent with the condition under the true data distribution,
rather than merely judging the marginal plausibility of the
candidate in isolation.

\subsection{Optimization algorithm}
\label{sec:cgan-opt}

The generator and discriminator are trained by alternating
stochastic gradient descent. At each training iteration the networks
see the complete measurement vector, never a partial record: a batch
element is either a fresh shot-noise realisation of one record
(Sec.~\ref{sec:protocol}) or a record from a different state
(Sec.~\ref{sec:amortised}). The discriminator parameters $\phi$ are updated first by
taking one gradient step on the loss~(\ref{eq:LD}) while holding
the generator parameters $\theta$ fixed:
\begin{equation*}
\phi \leftarrow \phi - \eta \, \nabla_{\phi} L_D,
\end{equation*}
where $\eta > 0$ is the learning rate. The generator parameters are
then updated by taking one gradient step on the
loss~(\ref{eq:LG}) while holding the newly updated discriminator
parameters fixed:
\begin{equation*}
\theta \leftarrow \theta - \eta \, \nabla_{\theta} L_G.
\end{equation*}
In practice, the Adam optimizer with momentum parameters $\beta_1 =
0.5$, $\beta_2 = 0.999$ is employed for both networks, following the
standard heuristic for stabilising GAN training. 
The penalty weight $\lambda$ is fixed at unity unless stated otherwise;
Sec.~\ref{sec:protocol} shows that the outcome is insensitive to it over a
range in which the balance between the two loss terms spans three orders of
magnitude.
% empirically, a value of $\lambda \in [10, 100]$ provides a good balance between
%learning the measurement distribution and enforcing pointwise data fidelity. 
Training runs for a fixed number of iterations. The fidelity with
respect to the target state is recorded during training as a
diagnostic only; it never enters the loss and is never used to stop
the run, so no experiment sees the target state before evaluation. At
convergence, the Cholesky-constrained generator
outputs a density matrix $\hat{\rho}_{\theta}$ that satisfies the
physical axioms by construction and whose Pauli expectations closely
reproduce the measured data.

  Fidelity is used as the principal
reconstruction metric, defined by the Bures--Uhlmann expression
rooted in Uhlmann's transition probability between quantum states
\cite{ref39}.

\begin{equation*}
F(\rho,\sigma)=\left[\operatorname{Tr}\left(\sqrt{\sqrt{\rho}\,\sigma\,\sqrt{\rho}}\right)\right]^{2}.
\end{equation*}

\section{Results}\label{sec:results}

\subsection{What the training protocol governs}
\label{sec:protocol}

\begin{figure*}[t]
\centering
\includegraphics[width=\textwidth]{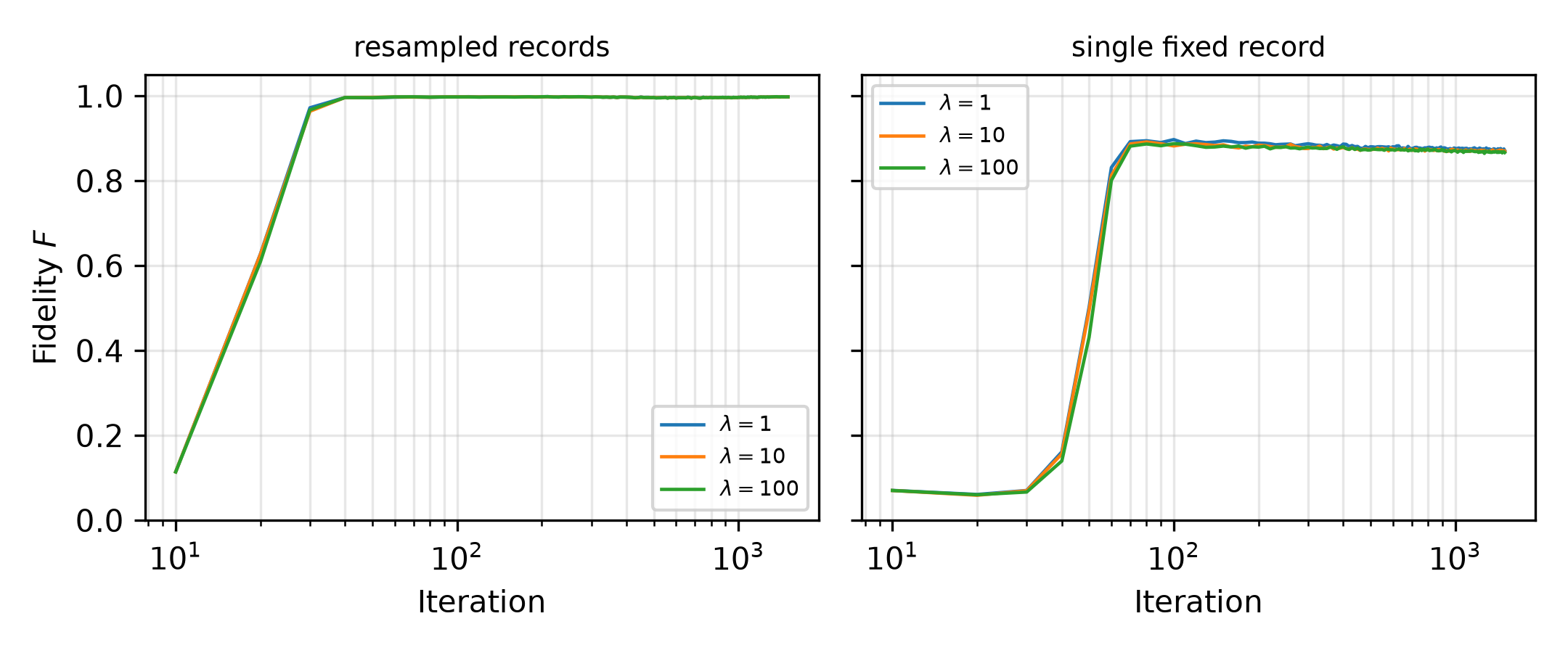}
\caption{Fidelity of the five-qubit GHZ reconstruction for
$\lambda=1,10,100$ (sum convention). Left: the discriminator is shown a fresh
shot-noise realisation of the record at every iteration; the three curves are
indistinguishable and all reach $F=0.9979$. Right: the same sweep when every
iteration reuses one fixed record; training stalls at $F=0.8717$ and the
weight again makes no difference. Logarithmic abscissa.}
\label{fig:protocol}
\end{figure*}

Reconstructing the five-qubit GHZ state~\cite{ref41}
$\frac{1}{\sqrt{2}}\left(|00000\rangle+|11111\rangle\right)$ is a demanding
test because the $32\times32$ density matrix contains only four theoretically
nonzero signal entries, two of which carry the nonlocal coherence between
$|00000\rangle$ and $|11111\rangle$. The remaining $1020$ entries should be
zero, and the $L^{1}$ penalty was introduced to suppress them.
Fig.~\ref{fig:protocol} reports what that penalty actually controls.

It does not control the dynamics. At $\lambda=1$ the balance between the data
and adversarial terms is $46{:}1$ in favour of the data term under the sum
convention and $1{:}22$ against it under the mean convention of
Eq.~(\ref{eq:LG}), a range of three orders of magnitude; the outcome is the
same either way (Table~\ref{tab:curves}). In the sum convention the three curves
$\lambda=1,10,100$ coincide, crossing $F=0.5$ at iteration $20$ and $F=0.99$ at
iteration $40$ and settling at $\CurvePoolX$. In the mean convention the result
is equally insensitive, with no systematic trend in $\lambda$. Adam shields the
dynamics from the gradient scale, which is why
the two conventions---whose data terms differ by a factor of $m$---behave the
same way.

What the reconstruction is sensitive to is the protocol---whether the
discriminator is shown a \emph{distribution} of measurement records or a single
one. If every iteration draws a fresh shot-noise realisation of the record, the
generator reaches $F=\CurvePoolX$. If every iteration reuses the same record,
which is the literal reading of the loss in Sec.~\ref{sec:method}, training
stalls at $F=\CurveSingleF$ for $\lambda=1$ and stays there
(Fig.~\ref{fig:protocol}, right).

The discriminator sees the condition and the candidate as two halves of one
input, so a real sample presents $(d^{\rm real},d^{\rm real})$ and a generated
one $(d^{\rm real},G(d^{\rm real}))$. The cheapest function that separates them
is ``are the two halves equal?'', which encodes nothing about the state; with a
fixed record that test can be sharpened onto a single point, the adversarial
gradient vanishes, and the generator improves only under the $L^1$ term.
Resampling moves the condition every batch,
forcing the discriminator to separate neighbourhoods rather than a point.
The weight makes no difference in either protocol.

All experiments that follow therefore resample the record at every iteration,
and $\lambda$ is fixed at unity rather than tuned: the sweep above shows no
systematic gain from tuning it.

\subsection{Thermal mixed states}

We consider thermal states of the transverse-field Ising model, a canonical many-body system. The Hamiltonian is given by
\[
H = -J \sum_{i=1}^{N-1} Z_i Z_{i+1} - h \sum_{i=1}^{N} X_i,
\]
with the number of qubits fixed at \(N=4\), the interaction strength \(J = 1.0\), and the transverse field \(h = 1.0\).

Thermal states extend the reconstruction problem beyond pure-state
tomography. A Gibbs state has the form $\rho_{\mathrm{th}} \propto
\exp(-\beta H)$, where $\beta$ is the inverse temperature. As the
temperature rises ($\beta$ decreases), the state degenerates toward
the isotropic maximally mixed state and the rank of the density
matrix grows markedly, making the reconstruction incrementally harder.
Three representative inverse temperatures are selected: $\beta = 2.0$
(low temperature), $\beta = 1.0$ (intermediate), and $\beta = 0.5$
(high temperature).

\begin{figure}[htbp]
\centering
\includegraphics[width=\columnwidth]{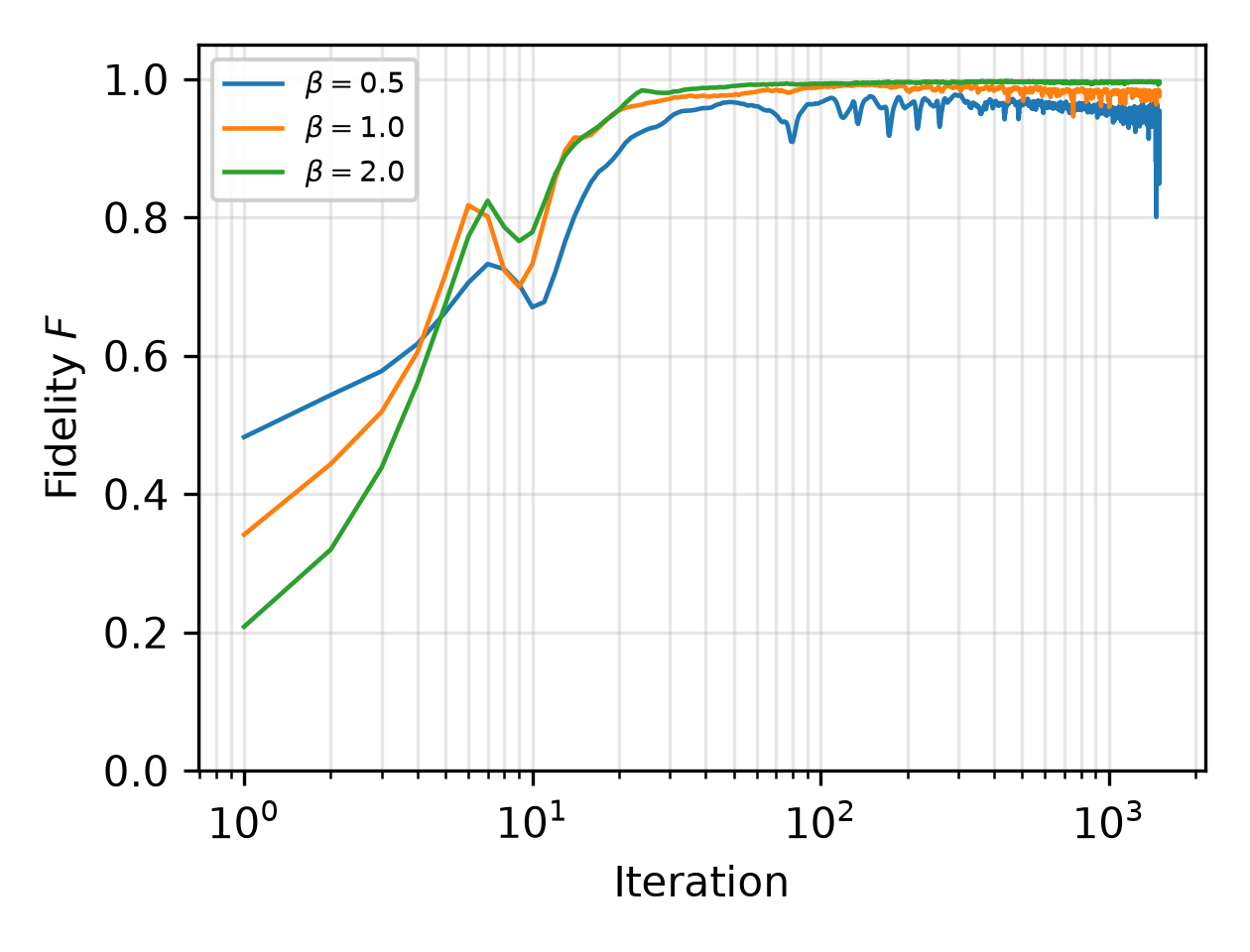}
\caption{Fidelity evolution of thermal Gibbs states at inverse
temperatures $\beta=0.5,1.0,2.0$. The high-temperature curve
($\beta=0.5$) starts highest, because a nearly maximally mixed state
is close to the untrained generator's output, and then jitters
between $0.8007$ and $0.9775$ instead of reaching unit fidelity; the lower
temperatures start from a random-guess fidelity and converge cleanly.}
\label{fig:thermal}
\end{figure}

As shown in Figure~\ref{fig:thermal}, the trajectories reproduce the
same temperature inversion at the iteration start: the high-temperature
state ($\beta = 0.5$) begins with the highest fidelity
($F = \CurveThInitLow$), whereas the low-temperature state
($\beta = 2.0$) begins very low ($F = \CurveThInitHigh$). The physical origin is that the high-entropy
state is close to uniform, as is the output of an untrained random
network, so the two are nearby in probability space; the
low-temperature state is highly ordered (approaching a pure state), so
a random guess is far off. The low- and intermediate-temperature
regimes ($\beta = 2.0$ and $\beta = 1.0$) then converge cleanly: with
thermal fluctuations suppressed and the state carrying definite
physical structure, the fidelity rises rapidly and stably, settling
at $F = \CurveThFinalLow$ and $F = \CurveThFinalMid$ respectively for the
remainder of the run.

The high-temperature state ($\beta = 0.5$) exposes a distinct
bottleneck. In the highly mixed regime the physical manifold becomes
extremely flat, and although the fidelity climbs quickly to a
moderate value it never settles on unit fidelity: after roughly
$10^{2}$ iterations the curve fluctuates persistently in the band
$0.8007$--$0.9775$, ending at $F = 0.9437$ rather than
converging. On this flat landscape the target expectations approach
uniformity, leaving the discriminator overly sensitive to minor
statistical deviations and producing sustained late-stage jitter
instead of a sharp breakdown or a clean plateau.

\begin{figure}[htbp]
\centering
\includegraphics[width=\columnwidth]{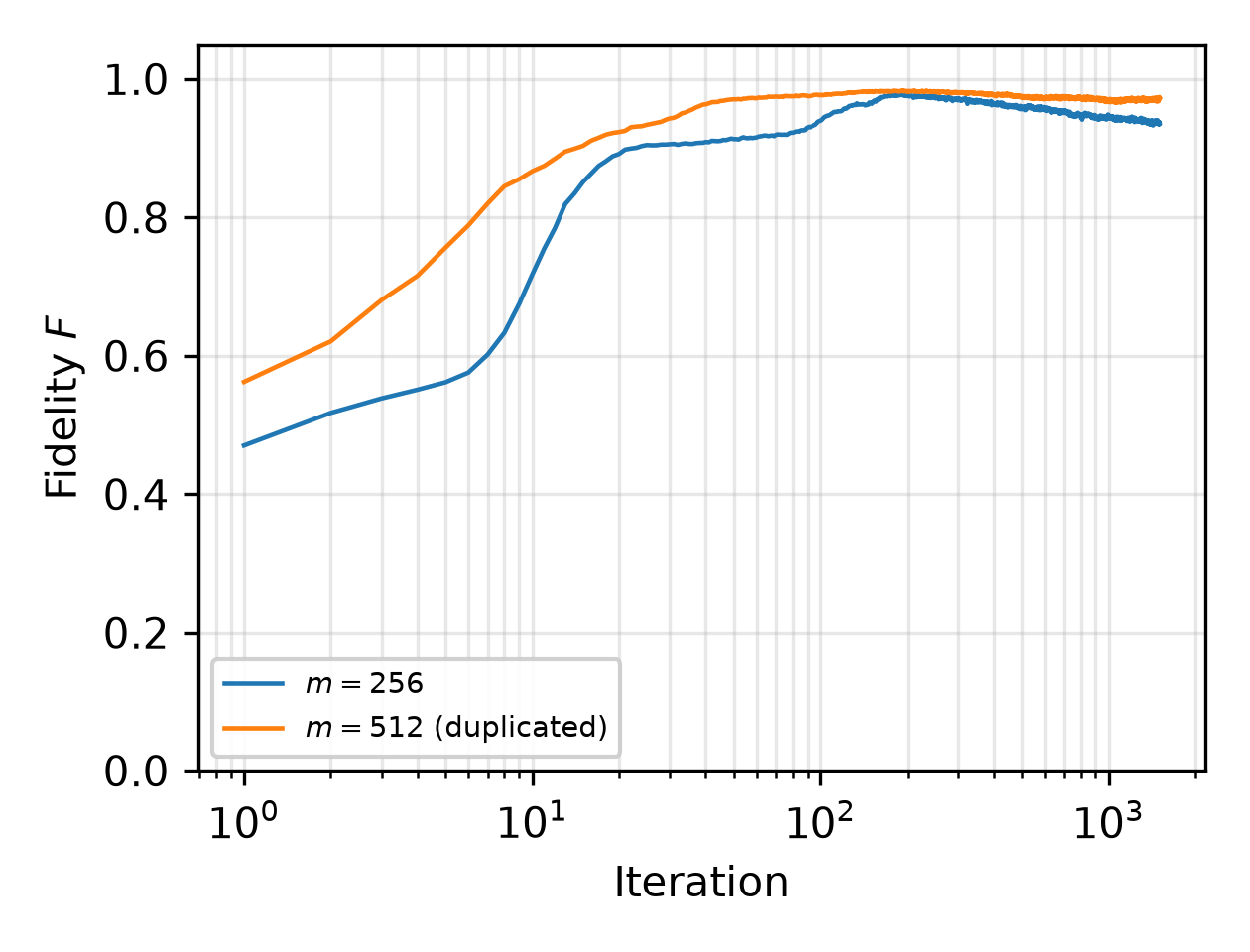}
\caption{Single run of the $\beta=0.5$ thermal reconstruction with the
complete basis ($m=256$) and with the duplicated conditioning vector
($m=512$). The controlled comparison, with ten seeds and matched
$L^1$ weight, is Table~\ref{tab:overcomplete}.}
\label{fig:oversample}
\end{figure}

To counteract this flat-landscape instability we duplicate the
conditioning vector by concatenating a copy of itself, expanding it from the $256$
complete Pauli observables to $512$ (see Figure~\ref{fig:oversample}).
The duplicated entry adds no physical information, yet the curve
stiffens: in the single run of Fig.~\ref{fig:oversample} the saturated
fidelity rises from $0.9363$ to $0.9726$, and the standard deviation of the
fidelity over the last $500$ iterations falls from $\CurveOverSdA$ to
$\CurveOverSdB$.
Section~\ref{sec:overcomplete} separates the contributions to
that gain with six configurations that change one ingredient at a time, and
shows that it comes from the
redundancy of the conditioning vector, not from additional
information and not from a change in the effective $L^1$ weight.
 
\subsection{Transverse-field Ising model ground states}

\begin{figure}[htbp]
\centering
\includegraphics[width=\columnwidth]{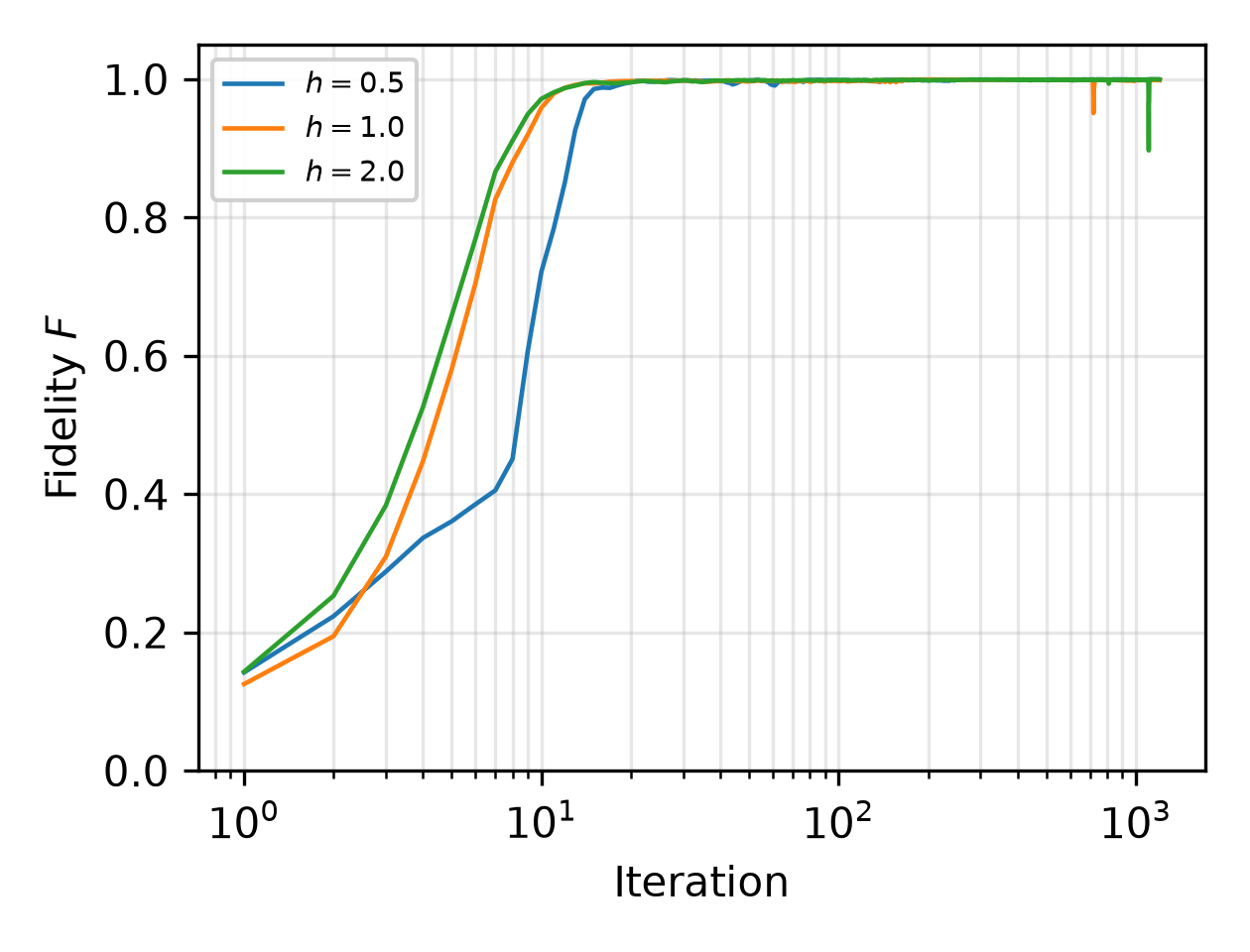}
\caption{Ground-state reconstruction fidelity of the 1D
transverse-field Ising model at $h=0.5,1.0,2.0$; all phases converge
above $F=0.9988$, with the ferromagnetic phase ($h=0.5$) showing an
initial lag.}
\label{fig:tfim}
\end{figure}

Figure~\ref{fig:tfim} displays the cross-phase convergence curves for the ground-state reconstruction of the one-dimensional transverse-field Ising model with \(J=1\) and three representative field strengths: \(h=0.5\) (ferromagnetic), \(h=1.0\) (critical), and \(h=2.0\) (paramagnetic). To assess the robustness of our method across distinct many-body phases, we target the exact ground-state density matrix for each case.

In all three cases, the reconstructed ground states attain fidelities exceeding \(0.9988\), confirming that the Cholesky-constrained generator faithfully captures the true quantum state. However, the convergence dynamics exhibit a clear phase dependence. Fig.~\ref{fig:tfim} uses a logarithmic abscissa; the paramagnetic (\(h=2.0\)) and critical (\(h=1.0\)) cases cross $F=0.5$ at iterations $4$ and $5$, whereas the ferromagnetic (\(h=0.5\)) case needs $9$---a factor of two that has narrowed by $F=0.99$ (iterations $18$ against $13$).

The ordering follows the structure of the target. At \(h=0.5\) the finite-size ground state is a near-degenerate cat superposition, and the generator has to establish coherence between two macroscopically distant configurations; its half-chain entropy is $0.865$~bits, against $0.411$ at the critical point and $0.126$ in the paramagnetic phase, and the iterations needed to cross $F=0.5$ fall in the same order (Table~\ref{tab:curves}). Despite these early differences, all three reconstructions complete the optimization within fewer than \(10^2\) iterations and stabilize above \(F=0.9988\), demonstrating that the Cholesky-based physical constraint is robust to gap closure and phase-transition singularities.

\begin{figure}[htbp]
\centering
\includegraphics[width=\columnwidth]{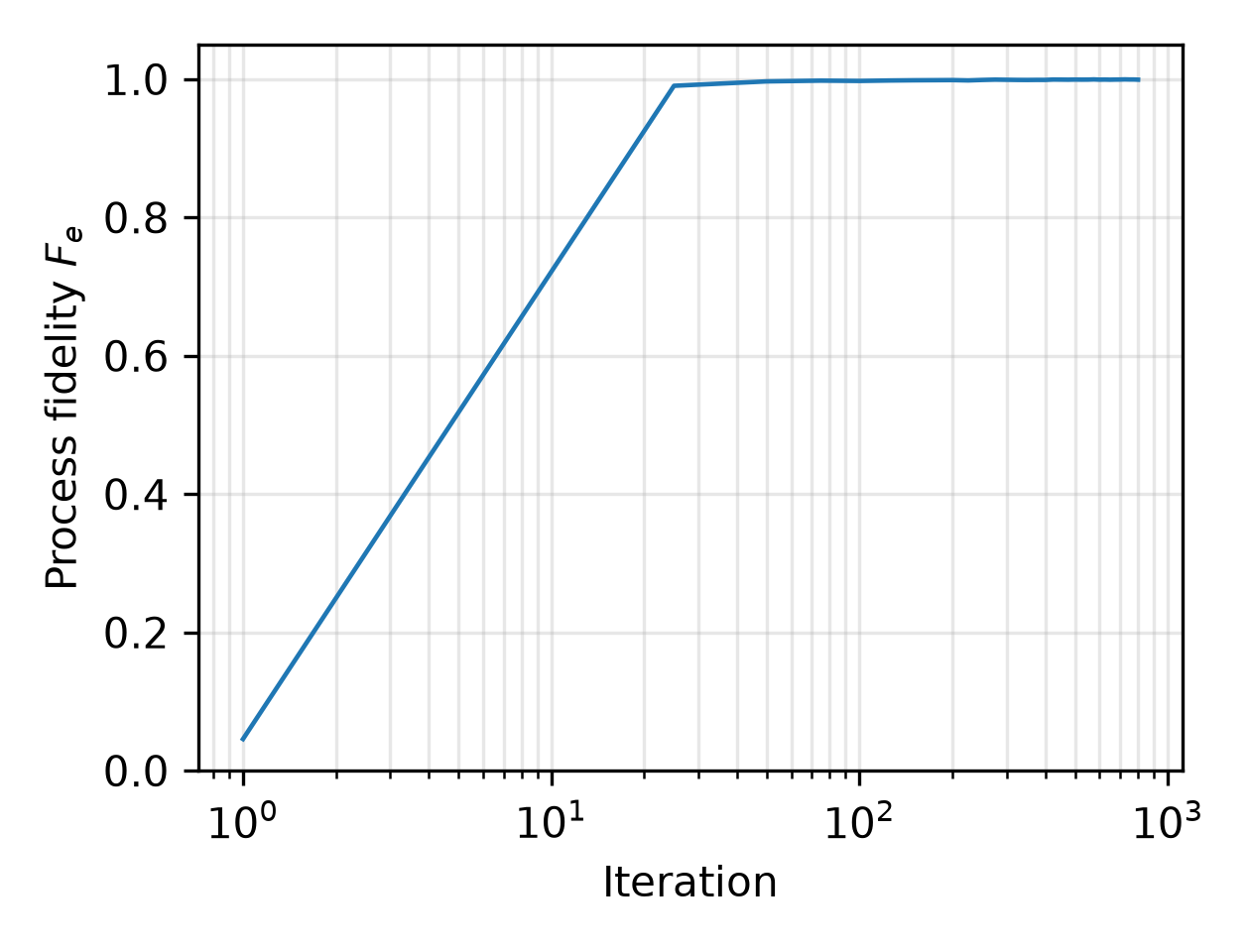}
\caption{Process fidelity of the reconstructed two-qubit CNOT channel
under the Choi-matrix generator; the run reaches $F_e=0.9993$ after
$800$ iterations.}
\label{fig:qpt-fidelity}
\end{figure}

\subsection{QPT process fidelity}

 The process-tomography experiment extends the
state-tomography architecture to quantum process tomography by
learning a physical channel rather than a density matrix. The
generator outputs Kraus operators for a two-qubit CNOT channel, and
the predicted measurement vector is obtained by applying the
generated channel to the $36$ probe states and evaluating the
expectation values of the $15$ non-identity two-qubit Pauli
observables. Specifically, the six single-qubit probe states are
\begin{equation*}
\mathcal{S} = \Big\{|0\rangle,\ |1\rangle,\ |+\rangle,\ |-\rangle,\
|+i\rangle,\ |-i\rangle\Big\},
\end{equation*}
where
\begin{equation*}
|\pm\rangle = \frac{|0\rangle \pm |1 \rangle}{\sqrt{2}}, \qquad |\pm
i\rangle = \frac{|0\rangle \pm i|1\rangle}{\sqrt{2}},
\end{equation*}
and the $36$ two-qubit probe states are given by all tensor products
\begin{equation*}
\rho^{(s)} = |\psi_a\rangle\!\langle\psi_a| \otimes
|\psi_b\rangle\!\langle\psi_b|, \qquad
|\psi_a\rangle,\,|\psi_b\rangle \in \mathcal{S},
\end{equation*}
The $15$ non-identity two-qubit Pauli observables are
\begin{equation*}
\mathcal{M} = \bigl\{\sigma_i \otimes \sigma_j \bigm|
(\sigma_i,\sigma_j) \in \{I,X,Y,Z\}^{2}
\setminus\{(I,I)\}\bigr\},
\end{equation*}
where $\sigma_0 = I,\ \sigma_1 = X,\ \sigma_2 = Y,\ \sigma_3 = Z$.
The full process measurement vector is therefore
\begin{equation*}
d^{\text{real}} = \Big(\operatorname{Tr}\!\big[(\sigma_i \otimes
\sigma_j)\,\mathcal{E}(\rho^{(s)})\big]\Big)_{s=1,\dots,36;\
(i,j)\neq(0,0)} \in \mathbb{R}^{540},
\end{equation*}
where $\mathcal{E}(\cdot)$ denotes the target quantum channel and
the data dimension is $36 \times 15 = 540$. Because the generator
output is converted into Kraus operators through the Cholesky
normalization layer, the reconstructed process remains completely
positive and trace preserving throughout training.

Figure~\ref{fig:qpt-fidelity} shows the resulting process-fidelity
curve. The fidelity leaves the random initialisation at $F_e=0.0458$
and reaches $0.9905$ within the first $25$ iterations: the $540$
measured expectations over-determine the $256$-dimensional Choi matrix,
so the first phase of training is a fast alignment that the linear
measurement map supports. After that the curve enters a slower
refinement regime in which the generator improves the remaining Choi
matrix agreement. The median over five seeds after $800$ iterations is
$F_e=\QptCnotChol$ (Table~\ref{tab:qpt}).

The shape of the curve is also informative. The early steep increase
suggests that the full measurement vector provides enough global
information for the network to identify the main CNOT structure,
while the later plateau reflects the harder task of matching the
remaining coherent process components.

\subsection{Comparison with classical estimators}
\label{sec:baselines}

The generator has so far been scored only against the target state. A
practitioner chooses between estimators, so we give the same measurement record
to the two standard classical reconstructions: linear inversion followed by the
Frobenius-nearest physical projection, which in the Pauli basis coincides
exactly with constrained least squares, and maximum likelihood for the Pauli
count data, computed with the standard $R\rho R$ fixed-point iteration.
Table~\ref{tab:baselines} reports medians over ten records at $10^3$ shots
per observable.

For pure and low-rank targets the ordering is unambiguous. On the five-qubit
GHZ state the projection reaches $F=\FGHZfiveProj$ and maximum likelihood
$F=\FGHZfiveMLE$, while the constrained generator reaches
$F=\FGHZfiveChol$. The constraint is doing real work rather than book-keeping:
the same architecture without the Cholesky layer returns matrices with negative
mass $\NGHZfiveRaw$ on five qubits and reaches only $F=\FGHZfiveRaw$, which the
post-hoc projection does not repair.

High-temperature mixed states behave differently. There the classical
estimators stall at $F=\FGibbsProj$ for projection and $\FGibbsMLE$ for
maximum likelihood, whereas the constrained generator reaches
$F=\FGibbsChol$, at the price of a wider run-to-run spread. We read this as
regularisation: the generator's low-dimensional parameterisation cannot chase
shot noise the way a maximum-likelihood estimate of $D^2-1$ free parameters
can, and on a nearly maximally mixed state that bias is worth more than the
variance it costs.

\begin{table*}[t]
\caption{Reconstruction fidelity for one measurement record per run (1000 shots per Pauli observable, 10 seeds). Interquartile range in brackets. ``Neg.\ mass'' is $\sum_{\lambda_i<0}|\lambda_i|$ of the raw estimator, i.e.\ evaluated before any post-hoc projection.}
\label{tab:baselines}
\begin{ruledtabular}
\begin{tabular}{lcccc}
State & Method & Median $F$ [IQR] & Neg.\ mass & Time (s) \\
\hline
$n=4$ GHZ & Linear inversion + projection & 0.8359 [0.8321, 0.8434] & 4.87e-17 & 0.01 \\
 & Maximum likelihood & 0.9995 [0.9995, 0.9996] & 9.3e-17 & 0.06 \\
 & Unconstrained, projected after & 0.8896 [0.8859, 0.8950] & 0.13 & 7.48 \\
 & Constrained, Cholesky (ours) & 0.9993 [0.9991, 0.9994] & 1.82e-13 & 10.66 \\
\hline
$n=5$ GHZ & Linear inversion + projection & 0.7115 [0.7088, 0.7168] & 8.22e-17 & 0.28 \\
 & Maximum likelihood & 0.9995 [0.9995, 0.9996] & 7.35e-17 & 6.24 \\
 & Unconstrained, projected after & 0.5940 [0.5498, 0.6269] & 0.721 & 15.28 \\
 & Constrained, Cholesky (ours) & 0.9995 [0.9991, 0.9995] & 1.49e-12 & 15.90 \\
\hline
$n=4$ Gibbs, $\beta=0.5$ & Linear inversion + projection & 0.9040 [0.9015, 0.9048] & 2.18e-17 & 0.00 \\
 & Maximum likelihood & 0.9040 [0.9023, 0.9054] & 1.52e-16 & 0.48 \\
 & Unconstrained, projected after & 0.9385 [0.9251, 0.9402] & 0.038 & 7.64 \\
 & Constrained, Cholesky (ours) & 0.9644 [0.9587, 0.9712] & 2.92e-10 & 8.19 \\
\hline
\end{tabular}
\end{ruledtabular}
\end{table*}

\subsection{Amortised reconstruction over a family of states}
\label{sec:amortised}

The experiments above train one generator per target state, which does not test
the amortisation that motivates a generative approach in the first place. We
therefore train a \emph{single} generator on $10^3$ ground states of the
one-dimensional transverse-field Ising model with $h$ uniform on $[0.2,2.5]$,
and evaluate it, without any retraining, on $200$ held-out states drawn from the
same family.

Table~\ref{tab:amortised} summarises the outcome. The amortised model
reconstructs held-out states at $F=\FAmortLone$ in $\TAmort$~ms per state,
about $\SAmort\times$ faster than per-state maximum likelihood
($F=\FAmortMLE$), and more accurate than it. The advantage is not a
speed-accuracy trade-off but a change of regime: the expensive part of the
problem is paid once, offline. Timing covers reconstruction only---the forward
pass for the amortised model, the optimisation for maximum likelihood. The
fidelity computation is a benchmark metric rather than part of either
estimator, so it is excluded from both.

Two qualifications matter. First, replacing the adversarial term with plain
$L^1$ regression does not cost accuracy: regression is at least as good
($F=\FAmortLone$ against $\FAmortCgan$, with interquartile ranges that do not
overlap) and trains faster, so the discriminator is not what makes
amortisation work. Second, the advantage is family dependent. On unstructured
random mixed states (Table~\ref{tab:amortised}, lower block) the
regression-trained model holds at $F=\FAmortRandLone$, level with per-state maximum
likelihood at $\FAmortRandMLE$, while the adversarial objective falls to
$\FAmortRandCgan$, below the first quartile of
both alternatives; when the family offers no structure
to exploit, the discriminator costs accuracy instead of buying it.

\begin{table*}[t]
\caption{Amortised reconstruction for $n=4$ and $10^3$ shots, evaluated on 200 held-out states that were never seen during training. Two state families are used: transverse-field Ising ground states with $h$ uniform on $[0.2,2.5]$, and unstructured random mixed states. ``Time'' is per-state wall clock on a CPU; the speed-up is relative to per-state maximum likelihood within the same block.}
\label{tab:amortised}
\begin{ruledtabular}
\begin{tabular}{lccc}
Method & Median $F$ [IQR] & Time per state & Speed-up \\
\hline
\multicolumn{4}{l}{\emph{TFIM ground states, $h\in[0.2,2.5]$}} \\
Amortised, $L^1$ only & 0.9983 [0.9980, 0.9986] & 0.023\,ms & $27005\times$ \\
Amortised, adversarial loss & 0.9959 [0.9940, 0.9972] & 0.042\,ms & $14827\times$ \\
Maximum likelihood (per state) & 0.9911 [0.9867, 0.9948] & 619\,ms & -- \\
Linear inversion + projection & 0.8412 [0.8369, 0.8459] & 0.445\,ms & -- \\
\hline
\multicolumn{4}{l}{\emph{Unstructured random mixed states}} \\
Amortised, $L^1$ only & 0.8934 [0.8711, 0.9091] & 0.021\,ms & $28035\times$ \\
Amortised, adversarial loss & 0.8471 [0.8144, 0.8713] & 0.022\,ms & $26897\times$ \\
Maximum likelihood (per state) & 0.8820 [0.8639, 0.9158] & 589\,ms & -- \\
Linear inversion + projection & 0.8739 [0.8628, 0.8855] & 0.440\,ms & -- \\
\hline
\end{tabular}
\end{ruledtabular}
\end{table*}

\subsection{What the over-complete measurement actually changes}
\label{sec:overcomplete}

The thermal-state analysis above reported that duplicating the $256$ Pauli
expectations into a $512$-dimensional conditioning vector suppresses the
late-stage fluctuations of the $\beta=0.5$ reconstruction, and attributed the
gain to redundant anchors stabilising a flat loss surface. That attribution is
too quick, because duplication also changes the effective weight of the $L^1$
term whenever the penalty is a sum over entries.
Table~\ref{tab:overcomplete} separates the two effects with six configurations
that differ in one ingredient at a time, all using the sum convention
of the original experiment.

Duplication raises the final fidelity from $\FOvrBase$ to $\FOvrDup$ and reduces
the late-stage jitter from $\JOvrBase$ to $\JOvrDup$, so the effect is real. It
is not a weight effect: halving $\lambda$ while keeping the duplicated input
leaves the result unchanged ($\FOvrMatch$), and doubling $\lambda$ at $m=256$
moves nothing ($\FOvrLamTwo$). Nor is it an information effect: an equal number
of genuinely independent shots reaches $\FOvrShots$ and a bootstrap draw of the
same size reaches $\FOvrBoot$. The gain therefore comes from the redundant
\emph{representation} of the conditioning vector, which averages each residual
twice and damps the discriminator's response to individual entries, not from
any new physical content. These numbers were obtained with the sum convention, in which duplication
changes the representation and the effective weight at once.

\begin{table*}[t]
\caption{What over-complete sampling actually changes. A high-temperature Gibbs state ($n=4$, $\beta=0.5$) is reconstructed with the sum convention of the original experiment; ten seeds per row. The last column is the standard deviation of the fidelity over the final three evaluations, a proxy for late-stage jitter.}
\label{tab:overcomplete}
\begin{ruledtabular}
\begin{tabular}{llcccc}
Configuration & $m$ & $\lambda$ & Shots & Best $F$ & Final $F$ (jitter) \\
\hline
complete basis, $m=256$ & 256 & 1.0 & 1000 & 0.9762 & 0.9500 (0.0014) \\
duplicated, $m=512$ (as originally proposed) & 512 & 1.0 & 1000 & 0.9821 & 0.9753 (0.0010) \\
duplicated, $m=512$, $\lambda$ halved & 512 & 0.5 & 1000 & 0.9830 & 0.9794 (0.0015) \\
$m=256$, $\lambda$ doubled & 256 & 2.0 & 1000 & 0.9754 & 0.9510 (0.0011) \\
$m=256$, twice the shots & 256 & 1.0 & 2000 & 0.9783 & 0.9708 (0.0015) \\
bootstrap, $m=512$ & 512 & 1.0 & 1000 & 0.9767 & 0.9725 (0.0013) \\
\hline
\end{tabular}
\end{ruledtabular}
\end{table*}

\subsection{Scaling to six qubits}
\label{sec:scaling}

\begin{figure}[t]
\centering
\includegraphics[width=\columnwidth]{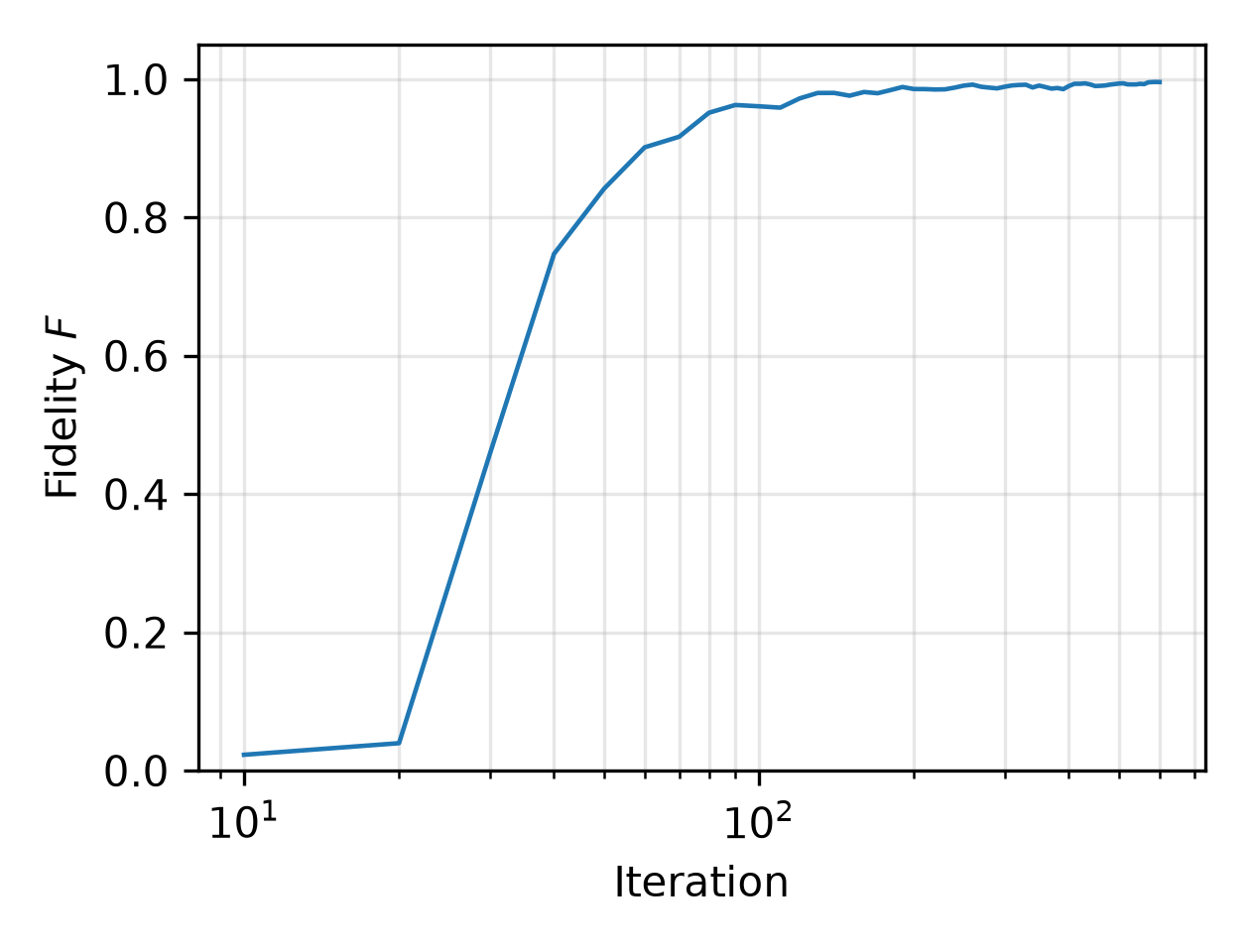}
\caption{Fidelity of the six-qubit GHZ reconstruction versus iteration on a
logarithmic abscissa ($2\times10^3$ shots per observable, $600$ iterations,
$16$~threads). The curve rises from the uninformative initial value and
approaches unity by iteration $200$, consistent with the $n=6$
row of Table~\ref{tab:scaling}.}
\label{fig:n6}
\end{figure}

The exponential cost is the structural limit of the approach, so it is worth
quantifying rather than asserting. Table~\ref{tab:scaling} reports three
training runs with identical settings, and Fig.~\ref{fig:n6} shows the $n=6$
trajectory. All three converge: $F=\ScaleFourFid$ at $n=4$, $\ScaleFiveFid$ at
$n=5$ and \ScaleSixFid{} at $n=6$. The price grows roughly as the algebra
predicts, though not uniformly over this
range: the per-iteration time rises from $\ScaleFourTime$~s to $\ScaleFiveTime$~s
and then to \ScaleSixTime~s, factors of two and four, and the peak working set
goes from $\ScaleFourMem$~MB to $\ScaleFiveMem$~MB and then to \ScaleSixMem~MB.
The generator parameters, $\ScaleFourPar$, $\ScaleFivePar$ and $\ScaleSixPar$,
approach the fourfold growth per qubit implied by the
$2\cdot4^{n}$ output.

These numbers locate the wall precisely. It is not the optimiser but the
representation: the $4^{n}$ input and the $2\cdot4^{n}$ output of the generator
multiply by four per qubit, so the output alone reaches $2\times10^{6}$ at
$n=10$ and $3\times10^{7}$ at $n=12$. A tensor-network generator, discussed in
the conclusion, is the natural way to move the constant rather than the
exponent.

\begin{table*}[t]
\caption{Scaling of the dense generator. Every row is one training run of 600 iterations on a GHZ target with $2\times10^3$ shots per Pauli observable, on a single CPU. Peak working set is measured for the run in isolation; parameters are those of the generator alone.}
\label{tab:scaling}
\begin{ruledtabular}
\begin{tabular}{cccccccc}
$n$ & $D=2^n$ & $m=4^n$ & Params & Time/iter (s) & Peak mem.\ (MB) & Total (s) & $F$ \\
\hline
4 & 16 & 256 & 263\,168 & 0.0169 & 100 & 10.1 & 0.9996 \\
5 & 32 & 1024 & 854\,528 & 0.0332 & 134 & 19.9 & 0.9999 \\
6 & 64 & 4096 & 3\,219\,968 & 0.1413 & 348 & 84.8 & 0.9947 \\
\hline
\end{tabular}
\end{ruledtabular}
\end{table*}

\subsection{Process tomography against a classical baseline}
\label{sec:qpt-baseline}

The process result quoted above had no comparison point. Table~\ref{tab:qpt}
supplies one, including two incoherently noisy channels. On the ideal CNOT the
constrained generator is the best of the three estimators
($F_e=\QptCnotChol$ against $\QptCnotLi$ for linear inversion followed by a
completely-positive projection), and the unconstrained variant fails badly,
returning Choi matrices whose negative eigenvalues sum to $\QptNegCnot$ against a
unit trace.

The ordering reverses once the noise is incoherent. At depolarising strengths
$p=0.05$ and $0.10$ the generator reaches $\QptDepFiveChol$ and $\QptDepTenChol$,
while linear inversion reaches $\QptDepFiveLi$ and $\QptDepTenLi$. The cause is
visible in the reconstruction itself: trained on data from the depolarised
channel, the generator returns a near-pure Choi matrix, with effective rank
$1.00$ and a fidelity of $0.9991$ to the ideal CNOT. The $\QptDepTenChol$ it
scores against the depolarised target is therefore barely above the $0.9063$
that reporting the ideal CNOT would give, whereas the depolarising noise spreads
the Choi spectrum
to an effective rank of $1.22$ and the generator does not follow it. Raising the
iteration count to $800$ and sweeping $\lambda$ over a decade moves the result by
less than $0.1$ percentage points, so this is a structural limitation of the
model rather than a tuning issue.

\begin{table*}[t]
\caption{Process tomography of a CNOT with and without incoherent noise ($2\times10^3$ shots, five seeds, median). ``Neg.\ mass'' refers to the Choi matrix before any projection, normalised to unit trace; a physical channel has zero. Process fidelity is evaluated on the normalized Choi matrices.}
\label{tab:qpt}
\begin{ruledtabular}
\begin{tabular}{llccc}
Channel & Method & Median $F_e$ & Neg.\ mass & Time (s) \\
\hline
CNOT & Linear inversion + CP projection & 0.9538 & 0.0485 & 0.6 \\
 & Unconstrained, projected after & 0.6263 & 1.96 & 6.3 \\
 & Constrained, Cholesky (ours) & 0.9990 & 1.07e-12 & 6.5 \\
\hline
CNOT + depol.\ $p=0.05$ & Linear inversion + CP projection & 0.9662 & 0.0302 & 0.7 \\
 & Unconstrained, projected after & 0.7390 & 6.14 & 6.5 \\
 & Constrained, Cholesky (ours) & 0.9564 & 4.16e-12 & 6.6 \\
\hline
CNOT + depol.\ $p=0.10$ & Linear inversion + CP projection & 0.9602 & 0.0165 & 0.7 \\
 & Unconstrained, projected after & 0.8350 & 4.28 & 7.2 \\
 & Constrained, Cholesky (ours) & 0.9099 & 1.73e-11 & 7.1 \\
\hline
\end{tabular}
\end{ruledtabular}
\end{table*}

\begin{table*}[t]
\caption{Readings taken off the learning curves of Figs.~1--5, listed here so that every number quoted in the text can be checked against a table rather than estimated from a plot. All entries are single runs; the multi-seed statistics are in Tables~I--III.}
\label{tab:curves}
\begin{ruledtabular}
\begin{tabular}{lll}
Figure & Quantity & Value \\
\hline
Fig.~1 (pooled) & $F=0.5$ / $F=0.99$ crossing, $\lambda=1,10,100$ & 20 / 40 (identical) \\
 & final $F$ & 0.9979, 0.9978, 0.9979 \\
Fig.~1 (single) & $F=0.5$ crossing & 50, 60, 60 \\
 & final $F$ & 0.8717, 0.8700, 0.8667 \\
Fig.~1 & data/adversarial balance at $\lambda=1$ & $46{:}1$ (sum), $1{:}22$ (mean) \\
Fig.~2 & initial $F$, $\beta=0.5,1,2$ & 0.4824, 0.3413, 0.2081 \\
 & final $F$ & 0.9437, 0.9808, 0.9952 \\
 & $F$ range of $\beta=0.5$ after iteration 100 & 0.8007--0.9775 \\
Fig.~3 & final $F$, $m=256$ / $m=512$ & 0.9363, 0.9726 \\
 & s.d. over the last 500 iterations & 0.0066, 0.0020 \\
Fig.~4 & $F=0.5$ crossing, $h=0.5,1,2$ & 9, 5, 4 \\
 & $F=0.99$ crossing & 18, 13, 13 \\
 & final $F$ & 0.9995, 0.9988, 0.9996 \\
 & half-cut entropy (bits) & 0.865, 0.411, 0.126 \\
Fig.~5 & $F_e$ at iteration 1 / 25 / 800 & 0.0458, 0.9905, 0.9993 \\
Sec.~III I & reconstruction: effective rank, $F_e$ to ideal CNOT & 1.00, 0.9991 \\
 & depolarised channel ($p=0.10$): effective rank & 1.22 \\
 & $F_e$ between ideal CNOT and depolarised channel & 0.9063 \\
\hline
\end{tabular}
\end{ruledtabular}
\end{table*}

\subsection{Limitations}
\label{sec:limitations}

These results are delimited on four sides. The largest system reconstructed here
is $n=6$, and Table~\ref{tab:scaling} shows why: the dense $2^{n}\times2^{n}$
representation multiplies by four per qubit, and the tensor-network compression
that would be needed to go further is not implemented. All experiments are
numerical, so the behaviour of the generator under detector inefficiency,
crosstalk and drift remains untested. The per-state numbers in
Table~\ref{tab:baselines} rest on ten measurement records each, and the
generator's run-to-run spread on mixed states is larger than that of maximum
likelihood. Finally, the process extension wins on coherent channels but loses
to linear inversion once the channel is depolarised, because the generator
returns the ideal channel instead of following the noise
(Table~\ref{tab:qpt}); the Cholesky prior should therefore not be expected to
help for strongly decoherent devices.

\section{Conclusion}\label{sec:conclusion}

This article has presented a physics-constrained conditional
generative network for quantum state reconstruction.
Embedding a differentiable Cholesky density-matrix layer at the
generator output guarantees Hermiticity, positive semi-definiteness,
and trace normalization by construction, while conditioning on Pauli
expectation values replaces repeated constrained inversion with a
single forward pass of a learned, amortizable inverse map.

The numerical experiments confirm the viability of this design
across structurally distinct target states. For the five-qubit GHZ
state the $L^{1}$ weight turns out not to matter: the balance between
the data and adversarial terms can be pushed from one extreme to the
other without changing the trajectory, because Adam absorbs the scale.
What does matter is whether the
discriminator is shown a distribution of measurement records or a
single one; the single-record protocol stalls at $F=\CurveSingleF$
because the discriminator degenerates into a test of exact equality.
Once the map is amortised the adversarial term can be dropped
altogether without loss of accuracy (Table~\ref{tab:amortised}).
Thermal Gibbs states converge cleanly at low and intermediate
temperature, but the high-temperature case ($\beta=0.5$) fluctuates
persistently in the late stage instead of reaching unit fidelity;
duplicating the conditioning vector stiffens the reconstruction and
lifts the saturated fidelity from $F=0.9363$ to $F=0.9726$,
and the controlled comparison of Sec.~\ref{sec:overcomplete} shows
that this gain originates in the redundancy of the conditioning vector
rather than in additional information, and that it is not explained by
a change of the effective $L^1$ weight. Finally, the one-dimensional
transverse-field Ising model ground state was reconstructed above
$F=0.9988$ across the ferromagnetic, critical, and paramagnetic
regimes.

Many machine-learning models can fit measurement data, but fitting
observables does not automatically guarantee that the inferred
object is a valid density matrix. By mapping every generator output
through a Cholesky construction, the constrained generator avoids non-physical
outputs at the source rather than correcting them after generation.

The comparison with standard estimators puts this in perspective. At
both $n=4$ and $n=5$ the constrained generator matches maximum
likelihood to within one part in a thousand ($\FGHZfourChol$ versus
$\FGHZfourMLE$, and $\FGHZfiveChol$ versus $\FGHZfiveMLE$), whereas
the same network without the Cholesky layer
collapses to $\FGHZfiveRaw$ and returns matrices with negative mass
$\NGHZfiveRaw$. On a quasi-maximally-mixed Gibbs state the generator
is ahead of both classical estimators ($\FGibbsChol$ versus
$\FGibbsMLE$). Trained once on a family of states, it reconstructs
held-out members at $\FAmortLone$ in $\TAmort$~ms, roughly
$\SAmort\times$ faster than per-state maximum likelihood.

We also probe the method at a larger system size. For the six-qubit
GHZ state ($n=6$, a $64\times64$ density matrix) the identical
pipeline still converges, to $F=\ScaleSixFid$ in the runs of
Table~\ref{tab:scaling}, confirming that the learning procedure
scales beyond small systems without qualitative degradation.
Nevertheless, the dominant cost is structural rather
than algorithmic: the generator represents the full
$2^{n}\times2^{n}$ density matrix, so both the memory footprint and
the per-iteration computational cost grow as
$\mathcal{O}(2^{2n})$ with the qubit count $n$. Each optimization
step therefore becomes exponentially more expensive, and the $4^{n}$
Pauli observables entering the conditioning vector inflate the
input and discriminator dimensions identically. For $n=6$ this
remains comfortably tractable on a shared-memory node, but for
$n\gtrsim 10$ the exponential wall makes the dense
$2^{n}\times2^{n}$ representation the fundamental bottleneck. The
forward-update cost of the generator itself scales linearly in the
network width and the number of iterations, yet it is dominated by
the matrix operations of the Cholesky layer and of the operator
trace products used to compute Born-rule expectations, both of which
scale with the square of the Hilbert-space dimension. This suggests
that a purely dense generative model cannot indefinitely outrun the
curse of dimensionality, however efficient the loss landscape may
be.

Several extensions follow naturally from the proposed framework.
First, tensor-network compression could reduce the memory burden of
the Cholesky representation and make the method more viable for
larger qubit counts; because the Cholesky layer acts only through
matrix products, it can be wrapped around a matrix-product density
operator or projected entangled pair structure to cut the storage
and per-iteration cost from $\mathcal{O}(2^{2n})$ to roughly
$\mathcal{O}(n)$ for weakly entangled states, trading the general
dense representation for a scalable structured one. Second,
active-learning strategies could select measurement bases
adaptively, reducing sample complexity by focusing measurement
resources on the most informative observables. Third, the
generative model could be migrated toward quantum-native
architectures, such as quantum generative adversarial networks or
variational quantum circuits, allowing part of the reconstruction
map to operate directly in Hilbert space.

The same constraint extends to process tomography, where a Choi matrix replaces
the density matrix (Sec.~\ref{sec:qpt-baseline}). There it succeeds on coherent
channels---the reconstructed CNOT reaches $F_e=\QptCnotChol$ against
$\QptCnotLi$ for linear inversion---and fails on depolarised ones, where the
generator returns the ideal channel rather than following the noise and the
classical estimator wins. The constraint should therefore be read as a guarantee
of physicality rather than as a universal accuracy advantage; a
device-certification pipeline would combine it with a classical estimator for
strongly decoherent devices.

%The broader implication is that quantum machine learning becomes more credible for physical reconstruction when it incorporates the mathematical laws of quantum mechanics into the model itself. The Cholesky layer, soft sparsity regularization, and over-complete sampling strategy together point toward a design principle for future tomography methods: neural networks should not merely approximate quantum data; they should be architected so that every approximation remains a physically meaningful quantum state.

\begin{acknowledgments}
Daming Li was supported by National Key Research and Development
Program of China (No. 2024YFA1014104).

\end{acknowledgments}

\end{document}